**Boundary conditions at the interface of finite thickness between ferromagnetic and antiferromagnetic materials**

Oksana Busel[1,*], Oksana Gorobets[1], Yuri Gorobets[1,2]

[1]Faculty of Mathematics and Physics, National Technical University of Ukraine "Igor Sikorsky Kyiv Polytechnic Institute", Prospect Peremohy 37, Kyiv, 03056, Ukraine
[2]Institute of Magnetism NAS and MES of Ukraine, Vernadskiy Av., 36-b, Kyiv, 03142, Ukraine

*opbusel@gmail.com

**Abstract:** Systematic approach has been applied to obtain the boundary conditions for magnetization at an interface between ferromagnetic (FM) and antiferromagnetic (AFM) materials in the continuous medium approximation. Three order parameters are considered inside an interface of finite thickness magnetization $\mathbf{M}$ of FM, magnetizations of both sublattices $\mathbf{M}_1$ and $\mathbf{M}_2$ of AFM. The boundary conditions are defined in terms of some average properties of the FM/AFM interface. The interface has a finite thickness which is much less than spin wave length. This approach allowed to take into account the interface anisotropy, interface symmetric exchange coupling and interface coupling resulting from inversion symmetry breaking in the vicinity of the interface.

**Introduction**

Antiferromagnetic spintronics has emerged recently as a research area [1]. Neel considered AFMs as extremely interesting but useless and for the long time nobody has touched this topic [2]. Back in the days it was hard for the outside observer to visualize the magnetic structure of AFM, which was caused by the fact that magnetic sublattices are compensated in the ground state and it doesn't usually create scattering fields unlike FMs. AFMs requires strong magnetic fields for the transition into magnetized state.

But AFM have a number of advantages that make them very interesting nowadays. First, they allow to work within THz frequencies [3], which is much more rapid than the frequencies accessible in FMs. Second, ability to manipulate AFMs, namely to L-vector – Neel order, as with electric and spin currents has been shown recently [4]. It is possible to create L-vector excitation in the AFM layer through interface exchange interaction of neighboring FM layer, as well as manipulating L-vector with the help of this [5, 6] as it is illustrated in Fig. 1. Moreover, methods of detection of L-vector has been discovered which are rather simple, for example, on basis of anisotropic magnetoresistance effect [7]. Also, X-ray magnetic linear dichroism (XMLD) provides one of the few tools to measure AFM order [8]. As a result, there is a possibility as to manipulate L-vector without application of strong magnetic fields and to detect it. And this has become popular for devices in the fields of magnonics and spintronics [6].

The interface between FM and AFM materials is investigated intensively during last decades, both theoretically and experimentally. Number of papers is dedicated to research of boundary conditions on the interface between FM and various materials, for example, the interface of a FM layer and a non-magnetic metal [9], as well as at the FM/AFM interface [10]. In particular, the influence of interface properties on the phenomenon of exchange bias of the hysteresis loop in FM/AFM structures was discovered a long time ago [11-13]. However, at the present time, interest to these effects still persists in connection with practical applications. There are theoretical models of this effect, which consider the interfaces between FM and AFM as either uncompensated [13] or compensated [14]. Besides, FM/AFM interface attracts attention of researchers in view of the observed domain structures in the magnetic force microscopy (MFM) experiments supported by micromagnetic calculations and magneto-resistive measurements which confirmed difference of the magnetic states in these microstructures for both compensated and uncompensated cases [14]. Polarization-dependent XMLD spectro-microscopy has been presented that reveals the micromagnetic structure on both sides of a FM/AFM interface [15]. Remanent hysteresis loops, recorded for individual FM domains, show a local exchange bias. The alignment of the FM spins is determined, domain by domain, by the spin directions in the underlying AFM layer [15]. In any case, the quality of the interface influences on the magnetic parameters of FM/AFM layers. Besides, the investigation of magnetic ordering on the boundary of FM/AFM is important for development antiferromagnetic spintronics [1]. However, the latter type of boundary conditions is insufficiently investigated.

In this paper a systematic approach has been applied for deriving boundary conditions at an interface between FM and AFM materials which was used to obtain the boundary conditions for magnetization at an interface between two FM materials in the continuous medium approximation [16]. This approach allows one to take into account the finite thickness of the FM/AFM interface as the boundary conditions are defined in terms of some average properties of the interface.

**Boundary conditions between FM/AFM:**

In order to develop the antiferromagnetic spintronics, it is important to have boundary conditions on the interface of FM/AFM. Thus, the task of this work is to find the most general form of the boundary conditions between FM and two-sublattice AFM taking into account the fact that the interface, as a composite material with finite thickness $\delta$ which is much less than the length of the spin wave $\lambda_{sw}$ [16].

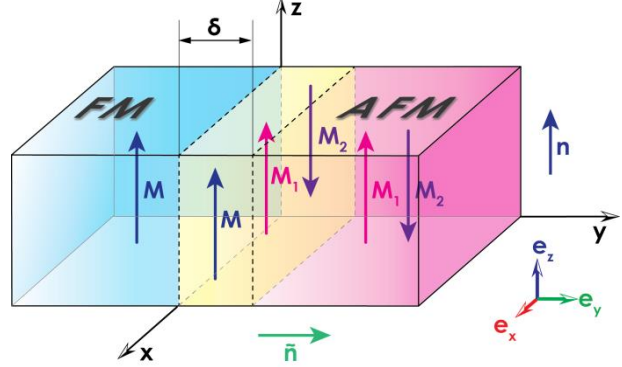

**Fig. 1.** The schematic image of the system consisting of FM, interface of finite thickness between FM/AFM and two-sublattice AFM

The normal to the interface of magnets is parallel to the *y*-axis. The general form of energy taking into account four energies, namely uniform and non-uniform exchange between sublattices, uniaxial magnetic anisotropy, antisymmetric (Dzyaloshinskii–Moriya) exchange interaction (DMI) and phenomenological non-local exchange coupling terms has the following form:

$$
\begin{aligned}
W = \int dy \Big\{ & A_{12}(y)\mathbf{M}_1\mathbf{M}_2 + A_1(y)\mathbf{M}_1\mathbf{M} + A_2(y)\mathbf{M}_2\mathbf{M} + \\
& + \alpha'_{12}(y)\left(\frac{\partial \mathbf{M}_1}{\partial y}\right)\left(\frac{\partial \mathbf{M}_2}{\partial y}\right) + \frac{1}{2}\alpha_1(y)\left(\frac{\partial \mathbf{M}_1}{\partial y}\right)^2 + \\
& + \frac{1}{2}\alpha_2(y)\left(\frac{\partial \mathbf{M}_2}{\partial y}\right)^2 + \alpha'_1(y)\left(\frac{\partial \mathbf{M}_1}{\partial y}\right)\left(\frac{\partial \mathbf{M}}{\partial y}\right) + \\
& + \alpha'_2(y)\left(\frac{\partial \mathbf{M}_2}{\partial y}\right)\left(\frac{\partial \mathbf{M}}{\partial y}\right) + \frac{1}{2}\alpha(y)\left(\frac{\partial \mathbf{M}}{\partial y}\right)^2 - \\
& - \frac{1}{2}\beta'_{12}(y)(\mathbf{Mn})^2 - \beta_1(y)(\mathbf{M}_1\mathbf{n})(\mathbf{Mn}) - \beta_2(y)(\mathbf{M}_2\mathbf{n})(\mathbf{Mn}) - \\
& - \beta_{12}(y)(\mathbf{M}_1\mathbf{n})(\mathbf{M}_2\mathbf{n}) - \frac{1}{2}\beta'_1(y)(\mathbf{M}_1\mathbf{n})^2 - \frac{1}{2}\beta'_2(y)(\mathbf{M}_2\mathbf{n})^2 + \\
& + d_{12}(y)[\mathbf{M}_1\times\mathbf{M}_2]\mathbf{n}' + d_1(y)[\mathbf{M}\times\mathbf{M}_1]\mathbf{n}' + d_2(y)[\mathbf{M}\times\mathbf{M}_2]\mathbf{n}' + \\
& + \sigma_{12}(y)\mathbf{M}_1\frac{\partial \mathbf{M}_2}{\partial y} + \sigma_1(y)\mathbf{M}_1\frac{\partial \mathbf{M}}{\partial y} + \sigma_2(y)\mathbf{M}_2\frac{\partial \mathbf{M}}{\partial y} + \\
& + \sigma_{21}(y)\mathbf{M}_2\frac{\partial \mathbf{M}_1}{\partial y} + \sigma'_1(y)\mathbf{M}\frac{\partial \mathbf{M}_1}{\partial y} + \sigma'_2(y)\mathbf{M}\frac{\partial \mathbf{M}_2}{\partial y} \Big\}
\end{aligned} \quad (1)
$$

Normal $\tilde{\mathbf{n}}$ is parallel to *y*-axis, unit vector of anisotropy $\mathbf{n}$ is parallel to the *z*-axis and unit vector of DMI is $\mathbf{n}'$. FM is magnetized along the *z*-axis: $\mathbf{M}$ is parallel to the *z*-axis (as shown in Fig. 1) and the *z*-axis is easy axis in the AFM, where the AFM vector is $\mathbf{L}=(\mathbf{M}_1-\mathbf{M}_2)$ (in the ground state); $A_{12}(y)$, $\alpha'_{12}(y)$ are uniform and non-uniform exchange parameters between 1st and 2nd AFM sublattices, respectively; $A_1(y)$, $A_2(y)$, $\alpha'_1(y)$, $\alpha'_2(y)$ are uniform and non-uniform exchange parameters between FM-1st and FM-2nd AFM sublattices, respectively; $\alpha(y)$, $\alpha_1(y)$, $\alpha_2(y)$ are non-uniform exchange parameters in the FM layer, 1st and 2nd AFM sublattices, respectively; $\beta_{12}(y)$, $\beta_1(y)$, $\beta_2(y)$ are uniaxial magnetic anisotropy parameters in the AFM layer, between FM-1st and FM-2nd AFM sublattices, respectively; $\beta_{12}'(y)$, $\beta_1'(y)$, $\beta_2'(y)$ are uniaxial magnetic anisotropy parameters in the FM layer, 1st and 2nd AFM sublattices, respectively; $d_{12}(y)$, $d_1(y)$, $d_2(y)$ are antisymmetric DMI parameters between AFM sublattices, FM-1st and FM-2nd AFM sublattices, respectively; $\sigma_{12}(y)$, $\sigma_{21}(y)$ are parameters of phenomenological non-local exchange coupling terms between 1st-2nd and 2nd-1st AFM sublattices, respectively; $\sigma_1(y)$, $\sigma_2(y)$, $\sigma_1'(y)$, $\sigma_2'(y)$ are parameters of phenomenological non-local exchange coupling terms between FM and 1st and 2nd AFM sublattices, respectively.

The constants in the energy (1) have typical dependency on the $y$ coordinate in the interface which is illustrated in Fig. 2:

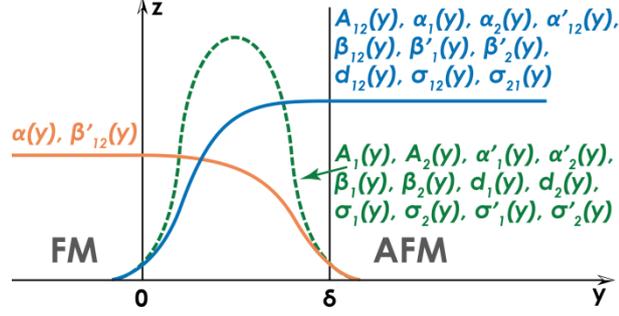

**Fig. 2.** The typical functional dependences of the constants in the energy (1) on the $y$ coordinate in the interface

Effective magnetic fields have been found:

$$\begin{cases} \mathbf{H}_{eff1} = -\delta W/\delta \mathbf{M}_1 \\ \mathbf{H}_{eff2} = -\delta W/\delta \mathbf{M}_2 \\ \mathbf{H}_{eff} = -\delta W/\delta \mathbf{M} \end{cases} \quad (2)$$

The Landau-Lifshitz equation is integrated over the thickness of the interface $y \in [0, \delta]$:

$$\int_0^\delta \partial \mathbf{M}_1/\partial t \, dy = g \int_0^\delta \left[ \mathbf{M}_1 \times \mathbf{H}_{eff1} \right] dy$$
$$\int_0^\delta \partial \mathbf{M}_2/\partial t \, dy = g \int_0^\delta \left[ \mathbf{M}_2 \times \mathbf{H}_{eff2} \right] dy . \quad (3)$$
$$\int_0^\delta \partial \mathbf{M}/\partial t \, dy = g \int_0^\delta \left[ \mathbf{M} \times \mathbf{H}_{eff} \right] dy$$

Oder parameters $\mathbf{M}$, $\mathbf{M}_1$, $\mathbf{M}_2$ are considered as slowly varying functions with $y \in [0, \delta]$, and the coefficients $A_1(y)$, $A_2(y)$, $A_{12}(y)$, $\alpha_1(y)$, $\alpha_2(y)$, $\alpha(y)$, $\alpha_1'(y)$, $\alpha_2'(y)$, $\alpha_{12}'(y)$, $d_{12}(y)$, $d_1(y)$, $d_2(y)$, $\beta_1(y)$, $\beta_2(y)$, $\beta_{12}(y)$, $\beta_1'(y)$, $\beta_2'(y)$, $\beta_{12}'(y)$, $\sigma_{12}(y)$, $\sigma_{21}(y)$, $\sigma_1(y)$, $\sigma_2(y)$, $\sigma_1'(y)$, $\sigma_2'(y)$ as rapidly varying functions (Fig. 2) for deriving boundary conditions taking into account $\delta \ll \lambda_{sw}$.

$$\int_0^\delta \left[ \mathbf{M}_1 \times \mathbf{H}_{eff1} \right] dy = 0$$
$$\int_0^\delta \left[ \mathbf{M}_2 \times \mathbf{H}_{eff2} \right] dy = 0 . \quad (4)$$
$$\int_0^\delta \left[ \mathbf{M} \times \mathbf{H}_{eff} \right] dy = 0$$

The notation is used, where the sign $\langle \rangle$ means averaging the function along the thickness of the interface.

The values of constants on the interface are presented in Table 1:

| | | | | |
|---|---|---|---|---|
| $\alpha_1(y)\|_0^\delta \equiv \alpha_1$ | $\int_0^\delta A_1(y)dy = \langle A_1 \rangle \delta$ | $\int_0^\delta \beta_1'(y)dy = \langle \beta_1' \rangle \delta$ | $\sigma_1'(y)\|_0^\delta \equiv 0$ | $\int_0^\delta \sigma_1'(y)dy = \langle \sigma_1' \rangle \delta$ |
| $\alpha_2(y)\|_0^\delta \equiv \alpha_2$ | $\int_0^\delta A_2(y)dy = \langle A_2 \rangle \delta$ | $\int_0^\delta \beta_2'(y)dy = \langle \beta_2' \rangle \delta$ | $\sigma_2'(y)\|_0^\delta \equiv 0$ | $\int_0^\delta \sigma_2'(y)dy = \langle \sigma_2' \rangle \delta$ |
| $\alpha(y)\|_0^\delta \equiv -\alpha$ | $\int_0^\delta A_{12}(y)dy = \langle A_{12} \rangle \delta$ | $\int_0^\delta d_1(y)dy = \langle d_1 \rangle \delta$ | $\sigma_1(y)\|_0^\delta \equiv 0$ | $\int_0^\delta \sigma_1(y)dy = \langle \sigma_1 \rangle \delta$ |
| $\alpha_1'(y)\|_0^\delta \equiv 0$ | $\int_0^\delta \beta_1(y)dy = \langle \beta_1 \rangle \delta$ | $\int_0^\delta d_2(y)dy = \langle d_2 \rangle \delta$ | $\sigma_2(y)\|_0^\delta \equiv 0$ | $\int_0^\delta \sigma_2(y)dy = \langle \sigma_2 \rangle \delta$ |
| $\alpha_2'(y)\|_0^\delta \equiv 0$ | $\int_0^\delta \beta_2(y)dy = \langle \beta_2 \rangle \delta$ | $\int_0^\delta d_{12}(y)dy = \langle d_{12} \rangle \delta$ | $\sigma_{12}(y)\|_0^\delta \equiv 0$ | $\int_0^\delta \sigma_{21}(y)dy = \langle \sigma_{21} \rangle \delta$ |
| $\alpha_{12}'(y)\|_0^\delta \equiv \alpha_{12}'$ | $\int_0^\delta \beta_{12}(y)dy = \langle \beta_{12} \rangle \delta$ | $\sigma_{21}(y)\|_0^\delta \equiv 0$ | $\int_0^\delta \sigma_{12}(y)dy = \langle \sigma_{12} \rangle \delta$ | $\langle \sigma_{12} \rangle = \langle \sigma_{21} \rangle$ |

**Table 1.** The notations of constants on the interface, where $\delta$ is the thickness of the interface

Then after integration of (4) the boundary conditions can be written in vector form as:

$$\begin{cases} \alpha_1 \left[ \mathbf{M}_1 \times \partial \mathbf{M}_1/\partial y \right] + \alpha_{12}' \left[ \mathbf{M}_1 \times \partial \mathbf{M}_2/\partial y \right] - \langle A_{12} \rangle \delta [\mathbf{M}_1 \times \mathbf{M}_2] - \langle A_1 \rangle \delta [\mathbf{M}_1 \times \mathbf{M}] + \\ + \langle \beta_{12} \rangle \delta [\mathbf{M}_1 \times (\mathbf{M}_2 \mathbf{n})\mathbf{n}] + \langle \beta_1 \rangle \delta [\mathbf{M}_1 \times (\mathbf{Mn})\mathbf{n}] + \langle \beta_1' \rangle \delta [\mathbf{M}_1 \times (\mathbf{M}_1 \mathbf{n})\mathbf{n}] - \\ - \langle d_{12} \rangle \delta [\mathbf{M}_1 \times [\mathbf{M}_2 \times \mathbf{n}']] - \langle d_1 \rangle \delta [\mathbf{M}_1 \times [\mathbf{n}' \times \mathbf{M}]] - \langle \sigma_1 \rangle \delta \left[ \mathbf{M}_1 \times \partial \mathbf{M}/\partial y \right] = 0 \\ \alpha_2 \left[ \mathbf{M}_2 \times \partial \mathbf{M}_2/\partial y \right] + \alpha_{12}' \left[ \mathbf{M}_2 \times \partial \mathbf{M}_1/\partial y \right] - \langle A_{12} \rangle \delta [\mathbf{M}_2 \times \mathbf{M}_1] - \langle A_2 \rangle \delta [\mathbf{M}_2 \times \mathbf{M}] + \\ + \langle \beta_{12} \rangle \delta [\mathbf{M}_2 \times (\mathbf{M}_1 \mathbf{n})\mathbf{n}] + \langle \beta_2 \rangle \delta [\mathbf{M}_2 \times (\mathbf{Mn})\mathbf{n}] + \langle \beta_2' \rangle \delta [\mathbf{M}_2 \times (\mathbf{M}_2 \mathbf{n})\mathbf{n}] - \\ - \langle d_{12} \rangle \delta [\mathbf{M}_2 \times [\mathbf{n}' \times \mathbf{M}_1]] - \langle d_2 \rangle \delta [\mathbf{M}_2 \times [\mathbf{n}' \times \mathbf{M}]] - \langle \sigma_2 \rangle \delta \left[ \mathbf{M}_2 \times \partial \mathbf{M}/\partial y \right] = 0 \\ -\alpha \left[ \mathbf{M} \times \partial \mathbf{M}/\partial y \right] - \langle A_1 \rangle \delta [\mathbf{M} \times \mathbf{M}_1] - \langle A_2 \rangle \delta [\mathbf{M} \times \mathbf{M}_2] + \langle \beta_{12}' \rangle \delta [\mathbf{M} \times (\mathbf{Mn})\mathbf{n}] \\ + \langle \beta_1 \rangle \delta [\mathbf{M} \times (\mathbf{M}_1 \mathbf{n})\mathbf{n}] + \langle \beta_2 \rangle \delta [\mathbf{M} \times (\mathbf{M}_2 \mathbf{n})\mathbf{n}] - \langle d_1 \rangle \delta [\mathbf{M} \times [\mathbf{M}_1 \times \mathbf{n}']] \\ - \langle d_2 \rangle \delta [\mathbf{M} \times [\mathbf{M}_2 \times \mathbf{n}']] - \langle \sigma_1' \rangle \delta \left[ \mathbf{M} \times \partial \mathbf{M}_1/\partial y \right] - \langle \sigma_2' \rangle \delta \left[ \mathbf{M} \times \partial \mathbf{M}_2/\partial y \right] = 0 \end{cases} \quad (5)$$

In the case of infinitely thin interface the constants of the uniform exchange between magnetizations of different magnetics have the form of $A_1(y) = A_1 \cdot \delta(y)$, $A_2(y) = A_2 \cdot \delta(y)$, where $\delta(y)$ is a delta function [17, 18]. The similar approach is a standard one for FM$_1$/FM$_2$ interface and in our case it implies that the coefficients of the uniform exchange are properly described with the formulas $\langle A_1 \rangle = A_1/\delta$, $\langle A_2 \rangle = A_2/\delta$ in the Table 1 for the interface of finite thickness, where $\delta$ is the interface thickness, $A_1$, $A_2$ are constants which may depend on interface properties, including interface roughness.

The boundary conditions in the vector form for the infinitely thin interface are sufficiently simplified since in this case $\delta \to 0$ and hence some terms containing $\delta$ will be equal to zero:

$$\begin{cases} \alpha_1 \left[ \mathbf{M}_1 \times \partial \mathbf{M}_1/\partial y \right] + \alpha_{12}' \left[ \mathbf{M}_1 \times \partial \mathbf{M}_2/\partial y \right] - A_1 [\mathbf{M}_1 \times \mathbf{M}] = 0 \\ \alpha_2 \left[ \mathbf{M}_2 \times \partial \mathbf{M}_2/\partial y \right] + \alpha_{12}' \left[ \mathbf{M}_2 \times \partial \mathbf{M}_1/\partial y \right] - A_2 [\mathbf{M}_2 \times \mathbf{M}] = 0 \\ \alpha \left[ \mathbf{M} \times \partial \mathbf{M}/\partial y \right] + A_1 [\mathbf{M} \times \mathbf{M}_1] + A_2 [\mathbf{M} \times \mathbf{M}_2] = 0 \end{cases} . \quad (6)$$

For example, the conditions (5) have been linearized taking into account the ground states of magnetization of FM, AFM and the interface, considering the small perturbations of order parameters relative to the ground state as follow:

$$\mathbf{M} = M_0 \mathbf{e}_z + \mathbf{m}, \quad \mathbf{m} = m_x \mathbf{e}_x + m_y \mathbf{e}_y$$
$$\mathbf{M}_1 = M_{01} \mathbf{e}_z + \mathbf{m}_1, \quad \mathbf{m}_1 = m_{1x} \mathbf{e}_x + m_{1y} \mathbf{e}_y \quad , \tag{7}$$
$$\mathbf{M}_2 = M_{02} \mathbf{e}_z + \mathbf{m}_2, \quad \mathbf{m}_2 = m_{2x} \mathbf{e}_x + m_{2y} \mathbf{e}_y$$

where

$$|\mathbf{m}| \ll M_0$$
$$|\mathbf{m}_1| \ll M_{01} \quad , \tag{8}$$
$$|\mathbf{m}_2| \ll M_{02}$$

then $\mathbf{m}$ is deviation magnetization of FM, $\mathbf{m}_1$, $\mathbf{m}_2$ are deviation magnetization of two sublattices AFM, $M_0$ is the projection of the magnetization of FM, $M_{01}$ and $M_{02}$ are projections of the magnetizations of the first and second sublattices respectively of the AFM to the $z$-axis in the ground state.

The notations are used $x, y = \upsilon$; $M_{02}/M_{01} = \gamma = -1$; $M_0/M_{01} = \gamma_0$; $M_0/M_{02} = \gamma_0/\gamma = -\gamma_0$.

The linearized boundary conditions have the following form:

$$\begin{cases} \langle A_{12} \rangle \delta(m_{1\upsilon} + m_{2\upsilon}) - \langle A_1 \rangle \delta(\gamma_0 m_{1\upsilon} - m_\upsilon) - \alpha_1 \partial m_{1\upsilon}/\partial y - \alpha'_{12} \partial m_{2\upsilon}/\partial y - \\ - \langle \beta_{12} \rangle \delta m_{1\upsilon} + \langle \beta_1 \rangle \delta \gamma_0 m_{1\upsilon} + \langle \beta'_1 \rangle \delta m_{1\upsilon} + \langle \sigma_1 \rangle \delta \partial m_\upsilon/\partial y = 0 \\ \langle A_{12} \rangle \delta(m_{1\upsilon} + m_{2\upsilon}) - \langle A_2 \rangle \delta(\gamma_0 m_{2\upsilon} + m_\upsilon) - \alpha_2 \partial m_{2\upsilon}/\partial y - \alpha'_{12} \partial m_{1\upsilon}/\partial y - \\ - \langle \beta_{12} \rangle \delta m_{2\upsilon} - \langle \beta_2 \rangle \delta \gamma_0 m_{2\upsilon} + \langle \beta'_2 \rangle \delta m_{2\upsilon} + \langle \sigma_2 \rangle \delta \partial m_\upsilon/\partial y = 0 \\ \langle A_1 \rangle \delta\left(m_{1\upsilon} - 1/\gamma_0 m_\upsilon\right) + \langle A_2 \rangle \delta\left(m_{2\upsilon} + 1/\gamma_0 m_\upsilon\right) + \alpha \partial m_\upsilon/\partial y + \langle \beta_1 \rangle \delta \, 1/\gamma_0 m_\upsilon - \\ - \langle \beta_2 \rangle \delta \, 1/\gamma_0 m_\upsilon + \langle \beta'_{12} \rangle \delta m_\upsilon + \langle \sigma'_1 \rangle \delta \partial m_{1\upsilon}/\partial y + \langle \sigma'_2 \rangle \delta \partial m_{2\upsilon}/\partial y = 0 \end{cases} \tag{9}$$

Then the linearized boundary conditions for the infinitely thin interface can be written as:

$$\begin{cases} A_1(\gamma_0 m_{1\upsilon} - m_\upsilon) + \alpha_1 \partial m_{1\upsilon}/\partial y + \alpha'_{12} \partial m_{2\upsilon}/\partial y = 0 \\ A_2(\gamma_0 m_{2\upsilon} - m_\upsilon) + \alpha_2 \partial m_{2\upsilon}/\partial y + \alpha'_{12} \partial m_{1\upsilon}/\partial y = 0 \\ A_1\left(m_{1\upsilon} - 1/\gamma_0 m_\upsilon\right) + A_2\left(m_{2\upsilon} + 1/\gamma_0 m_\upsilon\right) + \alpha \partial m_\upsilon/\partial y = 0 \end{cases} \tag{10}$$

The linearized boundary conditions in Eqs. (8) and (9) do not include DMI and are valid in the case when DMI gives a much smaller contribution than the exchange energy. The boundary conditions in the vector form (5) are also valid with considering the DMI.

The resulting boundary conditions on the FM/AFM interface are also useful for experimental researchers. Namely, the boundary conditions are applicable for the determination of the magnetization orientation in the interfacial boundary calculation for antiferromagnet-based tunnel junction, since it was demonstrated that efficient rotation of staggered moments in the antiferromagnet can be induced by the exchange-spring effect of the adjacent FM layer [19] and it was verified that antiferromagnetic moments in IrMn are persistently pinned along the easy direction of IrMn with in-plane fields due to the unidirectional anisotropy [20].

**Conclusions**

General form of boundary conditions has been found at the interface of the FM and two-sublattice AFM for the both interfaces of finite thickness and infinitely thin. The boundary conditions have general form and can be used for any type of FM/AFM interfaces for which the energy includes uniform and non-uniform exchange between sublattices, uniaxial magnetic anisotropy, DMI and phenomenological non-local exchange coupling terms in the form of Eq. (1). Thus, restrictions of generality in such boundary conditions are due to the assumption of isotropic exchange (the exchange-constant tensor has the diagonal form) and uniaxial magnetic anisotropy of FM and AFM. Herewith the interfacial roughness influences only the following coefficient values $\langle A_1 \rangle$, $\langle A_2 \rangle$, $\langle \beta_1 \rangle$, $\langle \beta_2 \rangle$, $\langle d_1 \rangle$, $\langle d_2 \rangle$, $\langle \sigma_1 \rangle$, $\langle \sigma_2 \rangle$, $\langle \sigma_1' \rangle$, $\langle \sigma_2' \rangle$ (see Table 1) for the interfaces of both finite thickness and infinitely thin.

The resulting boundary conditions are important for investigating the hysteresis loop of the FM/AFM structures, for studying the micromagnetic structure near the FM/AFM interface, for modeling the anisotropic magnetoresistance and propagation of spin waves in the FM/AFM system. The results obtained will contribute to a progress an area antiferromagnetic spintronics.

### Acknowledgement

This work was supported by the European Union's Horizon 2020 research and innovation programme under the Marie Sklodowska-Curie GA No. 644348 (MagIC).